\documentclass[12pt,a4paper]{article}
\usepackage{latexsym,amssymb,amsmath}

\newcommand{\neeg}{\overline{\phantom a}}
\newcommand{\wt}{\widetilde}

\begin{document}
\title{Upper bounds for the formula size of the majority function
\footnote{Research supported in part by RFBR, grants
11--01--00508, 11--01--00792, and OMN RAS ``Algebraic and
combinatorial methods of mathematical cybernetics and information
systems of new generation'' program (project ``Problems of optimal
synthesis of control systems'').}}
\date{}
\author{I. S. Sergeev}

\maketitle
\begin{abstract}
It is shown that the counting function of $n$ Boolean variables
can be implemented with the formulae of size $O(n^{3.06})$ over
the basis of all 2-input Boolean functions and of size
$O(n^{4.54})$ over the standard basis. The same bounds follow for
the complexity of any threshold symmetric function of $n$
variables and particularly for the majority function. Any bit of
the product of binary numbers of length $n$ can be computed by
formulae of size $O(n^{4.06})$ or $O(n^{5.54})$ depending on
basis. Incidental\-ly the bounds $O(n^{3.23})$ and $O(n^{4.82})$
on the formula size of any symmet\-ric function of $n$ variables
with respect to the basis are obtained.
\end{abstract}

\section{Introduction}

We consider the complexity of implementation of symmetric Boolean
functi\-ons with formulae over the basis $B_2$ of all binary
Boolean functions and over the standard basis $B_0 = \{ \wedge,
\vee,\neeg \}$. All necessary notions of formulae and complexity
$L_B(f)$ of implementation of  function $f$ with formulae over
basis $B$ one can find in~\cite{elu,eju}.

The best known results on the complexity and the depth of
implementa\-tion of symmetric functions with either circuits or
formulae over complete bases depend on the efficient
implementation of the counting function \linebreak
$C_n(x_1,\ldots,x_n)$ calculating the sum of Boolean variables
$x_1,\ldots,x_n$. Reduction to the compu\-ta\-tion of $C_n$ is a
way of minimization of the depth and the complexity of formulae
for multiplication of binary numbers.

In its turn, efficient circuits and formulae for the counting
function can be built of CSA-units.\footnote{CSA is abbreviature
for Carry Save Adder.} $(k,l)$-CSA of width 1 implements a Boolean
function $(x_1,\ldots,x_k) \to (y_1,\ldots,y_l)$ according to
condition $\sum a_i x_i = \sum b_j x_j$, where $k>l$ and constants
$a_i, b_j$ are the integer powers of two. $(k,l)$-CSA of arbitrary
width can be composed of parallel copies of width 1 CSA's. It
allows to reduce addition of $k$ numbers to addition of $l$
numbers.

With the use of appropriate CSA's and method~\cite{eppz} the
bounds $L_{B_2}(C_n)=O(n^{3.13})$, $L_{B_0}(C_n)=O(n^{4.57})$ were
obtained in~\cite{epz} improving upon preceding known results
$L_{B_2}(C_n)=O(n^{3.32})$~\cite{epe} and
$L_{B_0}(C_n)=O(n^{4.62})$~\cite{ekh} (some earlier results see
in~\cite{eppz}).

In the recent paper~\cite{edkky} it was built a new CSA (called
MDFA) which allows to implement the function $C_n$ with a circuit
over $B_2$ of the best known complexity $4.5n$ (improving the old
previous bound $5n$). Efficient use of the CSA~\cite{edkky}
implies encoding of some pairs of bits $u, v$ in the form
$(u\oplus v, v)$ (such encoding was introduced in~\cite{est}). A
benefit in complexity is due to (a) one can compute a pair
$(u\oplus v, v)$ not harder than $(u, v)$, (b) following
computations need $u\oplus v$ rather than $u$~--- that's why one
Boolean addition can be saved.

Note that by the above reasons one can expect MDFA to be efficient
for implementation with formulae. Indeed, MDFA allows shorter
formulae for $u \oplus v$ than formulae for $u$ and $v$ together.

This observation was exploited implicitly in the construction of
the formula efficient CSA in~\cite{epz}. The CSA contains MDFA
with its outputs connected to the standard $(3,2)$-CSA $FA_3$.
What makes the CSA~\cite{epz} more efficient than the standard
$(3,2)$-CSA is exactly intermediate $(u\oplus v, v)$ encoding (the
$(3,2)$-CSA alone leads to the bound $L_{B_2}(C_n)=O(n^{3.21})$,
see~\cite{eppz}).

Moreover, as far as MDFA saves circuit complexity, CSA~\cite{epz}
also does it: its circuit complexity is 14.\footnote{Formula for
the circuit implementation slightly differs from the shortest
one.} Thus, the CSA allows to implement $C_n$ with $(14/3)n$
circuit complexity. So, it was possible to overcome the $5n$
barrier in the 90-es.

It is natural to conclude that the CSA~\cite{epz} does not use
MDFA optimally to construct shorter formulae. We will show below
that one can obtain the bound $L_{B_2}(C_n)=O(n^{3.06})$ via more
independent way of exploiting MDFA. However, proposed method is
also not optimal.

An analogous idea works in the case of basis $B_0$. In the case
one can try monotone encoding $(uv,\;u\vee v)$ of a pair of bits
$u$ and $v$. Such encoding makes the most significant outputs of a
CSA to be monotone functions of inputs (in fact, these functions
are threshold if all inputs are of the same significance). It can
be reasonable since the known CSA's over $B_0$ have non-monotone
outputs to be most difficult for implementation.

We will describe below $(5,3)$-CSA SFA$_5$, an analogue of MDFA
for mono\-tone encoding. With the use of it the bound
$L_{B_0}(C_n)=O(n^{4.55})$ is rather simple to obtain. Slightly
better bound $L_{B_0}(C_n)=O(n^{4.54})$ follows from the more
complicated construction based on the $(7,3)$-CSA~\cite{ekh} and
monotone encoding of triples of bits.

Let $S_n$ denote the class of symmetric Boolean functions of $n$
variables. New upper bounds for majority function does not provide
automatic reduc\-tion of the size of formulae implementing
functions from $S_n$ due to the fact that the known
methods~\cite{ekh,eppz0} limit the efficiency of CSA's with inputs
and outputs of multiple types.

However, it is not hard to aggregate several CSA's with
non-standard encoding of bits into single CSA with the standard
encoding and apply the method~\cite{eppz0} to obtain bounds
$L_{B_2}(S_n)=O(n^{3.23})$, $L_{B_0}(S_n)=O(n^{4.82})$ improving
earlier results $L_{B_2}(S_n)=O(n^{3.30})$~\cite{eppz0},
$L_{B_2}(S_n)=O(n^{3.37})$~\cite{epe},
$L_{B_0}(S_n)=O(n^{4.85})$~\cite{eppz0},
$L_{B_0}(S_n)=O(n^{4.93})$~\cite{ekh}.

It is worth noting that basic CSA's in the present paper are
structurally similar or in any case are not more complicated than
the CSA's in preceding papers. So, the improvement in complexity
bounds is entirely a result of exploiting the idea of alternative
encoding of bits.

\section{Formulae over $B_2$}

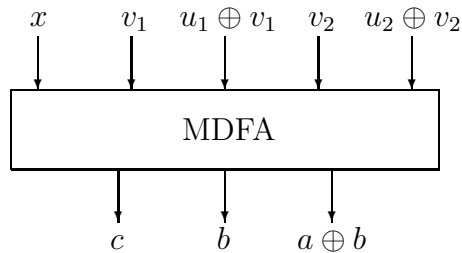
\begin{figure}[htb]
\begin{picture}(170,90)(-115,0)

\multiput(0,30)(160,0){2}{\line(0,1){30}}
\multiput(0,30)(0,30){2}{\line(1,0){160}}

\multiput(40,30)(40,0){3}{\vector(0,-1){20}}
\multiput(10,80)(35,0){5}{\vector(0,-1){20}}

\put(64,41){MDFA}

\put(7,84){$x$} \put(41,84){$v_1$} \put(63,84){$u_1\oplus v_1$}
\put(111,84){$v_2$} \put(132,84){$u_2\oplus v_2$}

\put(37,0){$c$} \put(77,0){$b$} \put(107,0){$a\oplus b$}

\end{picture}
\caption{MDFA block-diagram}
\end{figure}

Fig. 1 shows a block-diagram of MDFA. Functional definition and
formulae to compute outputs are given below.

$$ 2(a+b)+c = x+u_1+u_2+v_1+v_2. $$
\begin{equation}\label{emdfa}
\begin{array}{c}
\phantom{\displaystyle \sum}  c  = x \oplus (u_1 \oplus v_1)
\oplus (u_2 \oplus v_2), \qquad
    b  = (x\oplus v_1)(u_1 \oplus v_1) \oplus v_1, \phantom{\displaystyle \sum}  \\
  (a \oplus b) = ((x\oplus v_1) \vee (u_1 \oplus v_1)) \oplus
    (x\oplus (u_1 \oplus v_1) \oplus v_2)\overline{(u_2 \oplus
    v_2)}.
\end{array}
\end{equation}

To obtain $O(n^{3.06})$ complexity bound one can use CSA shown in
Fig. 2. It contains two isolated MDFA's, which are identical up to
encoding of a pair of inputs.

\begin{figure}[htb]
\begin{picture}(370,120)(-13,30)

%%% MDFA carcas

\put(0,30){
\begin{picture}(160,90) % MDFA
\multiput(0,30)(160,0){2}{\line(0,1){30}}
\multiput(0,30)(0,30){2}{\line(1,0){160}}
\multiput(40,30)(40,0){3}{\vector(0,-1){20}}
\put(10,80){\vector(0,-1){20}} \put(45,110){\vector(0,-1){50}}
\put(46.7,110){\vector(4,-3){29}} \put(80,110){\vector(0,-1){20}}
\put(80,85){\circle{10}} \put(80,80){\line(0,1){10}}
\put(75,85){\line(1,0){10}}
\multiput(80,80)(35,0){3}{\vector(0,-1){20}} \put(64,41){MDFA}
\end{picture}}

%%%  signatures

\put(11,114){$x_1$} \put(44,144){$x_2$} \put(80,144){$x_3$}
\put(115,114){$v_1$} \put(137,114){$u_1 \oplus v_1$}

\put(40,30){$c_1$} \put(81,30){$b_1$} \put(108,30){$a_1\oplus
b_1$}

%%% MDFA carcas

\put(195,45){
\begin{picture}(160,90) % MDFA
\multiput(0,30)(160,0){2}{\line(0,1){30}}
\multiput(0,30)(0,30){2}{\line(1,0){160}}
\multiput(40,30)(40,0){3}{\vector(0,-1){20}}
\multiput(10,80)(35,0){5}{\vector(0,-1){20}} \put(64,41){MDFA}
\end{picture}}

%%%  signatures

\put(204,129){$x_4$} \put(239,129){$v_2$} \put(262,129){$u_2
\oplus v_2$} \put(309,129){$v_3$} \put(332,129){$u_3 \oplus v_3$}

\put(234,45){$c_2$} \put(275,45){$b_2$} \put(302,45){$a_2\oplus
b_2$}

\end{picture}
\caption{The main CSA}
\end{figure}
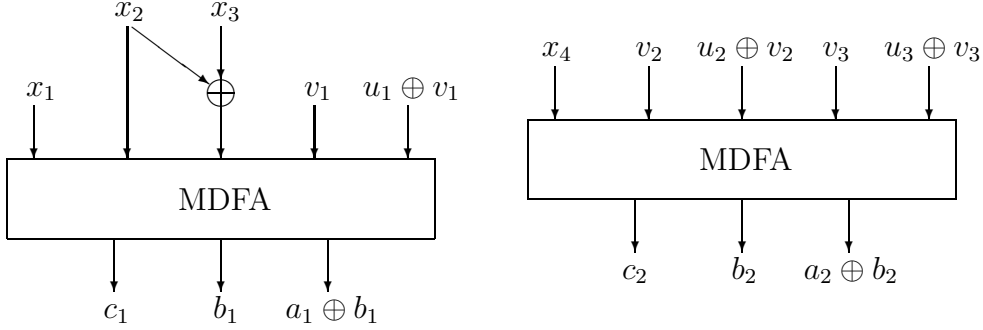

The CSA has inputs and outputs of two types: standard bits and
pairs of bits encoded as $(u\oplus v,v)$.

Consider the size of formulae implementing inputs and outputs of
the first type. Denote it by $X_i$ for inputs $x_i$ and by $C_i$
for outputs $c_i$. To deal with inputs and outputs of the second
type $(u\oplus v,v)$ introduce a quantity $\max \{ V,\,
U^+/\alpha\}$ where $V$, $U^+$ is the size (or an upper estimate
of the size) of formulae implementing $v$ and $u\oplus v$
respectively, $\alpha$ is a parameter to be chosen later. Denote
this quantity by $U_i$ for inputs $(u_i\oplus v_i,v_i)$ and by
$A_i$ for outputs $(a_i\oplus b_i,b_i)$.

According to~(\ref{emdfa}), the following inequalities hold:
\begin{equation}\label{esys}
\begin{array}{c}
C_1 \le X_1 + X_2 + X_3 + \alpha U_1, \\ \phantom{\displaystyle
\sum}
 C_2 \le X_4 + \alpha U_2 + \alpha U_3, \phantom{\displaystyle \sum}
 \\ \phantom{\displaystyle
\sum}
 A_1 \le \max \left\{ X_1 + X_2 + 3X_3, \;
\frac2{\alpha}X_1 + \frac2{\alpha}X_2 + \frac3{\alpha}X_3 +
\frac{\alpha+1}{\alpha}U_1 \right\}, \phantom{\displaystyle \sum}
\\ \phantom{\displaystyle \sum}
 A_2 \le \max \left\{ X_4 + (\alpha+2)U_2, \;
\frac2{\alpha}X_4 + \frac{2\alpha+1}{\alpha}U_2 +
\frac{\alpha+1}{\alpha}U_3 \right\}. \phantom{\displaystyle \sum}
\end{array}
\end{equation}

It follows from~\cite{eppz,epz} that if
\begin{equation}\label{einq}
\begin{array}{c}
X_1^p + X_2^p + X_3^p + X_4^p - C_1^p - C_2^p > 0, \\
\phantom{\displaystyle \sum}U_1^p + U_2^p + U_3^p - A_1^p - A_2^p
> 0, \phantom{\displaystyle \sum}
\end{array}
\end{equation}
for some $p>0$, some $X_i > 0$ and $U_i > 0$ (and also for some
$\alpha>0$) then $L_{B_2}(C_n) = O\left(n^{1/p+o(1)}\right)$, see
also Appendix. Use upper bounds~(\ref{esys}) instead of $C_i$ and
$A_i$ to check that inequalities~(\ref{einq}) hold when
$\alpha=2.906$, $p=0.327781$, $X_1=X_2=1$, $X_3=0.5149081$,
$X_4=1.9198088$, $U_1=1.2176395$, $U_2=1.0031176$,
$U_3=2.3573055$. Consequently, $L_{B_2}(C_n) = O(n^{3.0509})$.

To obtain tighter estimates of the efficiency of MDFA one can
split formally the second type of encoding into several types with
different values of $\alpha$ (say, uniformly distributed in some
segment) and consider a set of MDFA's with inputs and outputs of
all possible types. However, the implied calculation looks rather
laborious if not to exploit some additional considera\-tions. This
observation is already involved partly in the construction of fig.
2: ratio of sizes of formulae for $x_2\oplus x_3$ and $x_2$
differs from $\alpha$.

To estimate the complexity of a symmetric function consider the
following sequence of CSA's with standard encoding of bits. In the
proposed sequence the $m$-th CSA contains $m$ MDFA's connected in
a chain and an outer CSA $FA_3$ (outputs of each MDFA are
connected to the inputs $v_2$, $u_2\oplus v_2$ of the next MDFA in
notation of fig. 1). The first CSA in the sequence ($m=1$) is the
CSA from~\cite{epz}. For $m=4$ we get $(15,6)$-CSA, which, if
taken independently, allows to implement $C_n$ with the formula of
size $O(n^{3.089})$.\footnote{Vector of sizes of outputs of the
CSA can be produced from the vector of sizes of inputs via
multiplication by the matrix
$$ \left( \begin{matrix}
0 & 0 & 0 & 0 & 0  \\
0 & 0 & 0 & 0 & 0  \\
0 & 0 & 0 & 0 & 1    \\
0 & 1 & 1 & 1 & 2   \\
1 & 2 & 2 & 3 & 3   \\
1 & 4 & 4 & 9 & 3
\end{matrix}\quad\begin{matrix}
 0 & 0 & 0 & 0 & 0 & 1 & 1 & 1 & 1 & 1 \\
 0 & 0 & 1 & 1 & 1 & 1 & 2 & 2 & 2 & 3 \\
 1 & 1 & 2 & 2 & 3 & 1 & 2 & 3 & 3 & 6  \\
 2 & 3 & 3 & 3 & 6 & 1 & 2 & 3 & 3 & 6 \\
 3 & 6 & 3 & 3 & 6 & 1 & 2 & 3 & 3 & 6 \\
 3 & 6 & 3 & 3 & 6 & 1 & 2 & 3 & 3 & 6
\end{matrix}\right)$$ }

With the use of this $(15,6)$-CSA and method~\cite{eppz0} one can
implement a $k$-th significant bit of $C_n$ with complexity
$O(n^{2.2285}\cdot2^k)$. Thus, the bound $L_{B_2}(S_n) =
O(n^{3.2285})$ follows, see Appendix for proof.

\section{Formulae over $B_0$}

The present section includes two examples of CSA's which are
efficient for constructing formulae for $C_n$ over $B_0$. The
first CSA is shown in fig. 3. Functional definition of the basic
CSA SFA$_5$ (Sorting Full Adder) is similar to that of MDFA.
Outputs are implemented by formulae:
\begin{equation}\label{esfa5}
\begin{array}{c}
\phantom{\displaystyle \sum} c = (x_1 \oplus (u_1 \oplus v_1))
\oplus (u_2 \oplus v_2) =
 \psi \overline{\chi} \vee  \overline{\psi} \chi, \phantom{\displaystyle \sum} \\
%\phantom{\displaystyle \sum}
\psi =  x_1\left((u_1v_1) \vee \overline{(u_1 \vee v_1)}\right)
\vee \overline{x_1}\overline{(u_1v_1)}(u_1 \vee
 v_1), \;\,  \chi = \overline{(u_2v_2)}(u_2 \vee  v_2), %\phantom{\displaystyle \sum}
 \\
a_1b_1 = T_5^4(x_1,u_1,v_1,u_2,v_2) = \displaystyle\bigvee_{i+j=4}
T_3^i(x_1,u_1,v_1) T_2^j(u_2,v_2) =
\\ = (x_1(u_1 \vee v_1) \vee (u_1v_1))(u_2v_2) \vee x_1(u_1v_1)(u_2 \vee
v_2), \end{array}
\end{equation}
where $T_n^k$ is the threshold monotone function of $n$ variables
with the threshold $k$; output $a_1 \vee b_1 =
T_5^2(x_1,u_1,v_1,u_2,v_2)$ is a function dual to $a_1b_1$, so it
can be implemented with the dual formula.

\begin{figure}[htb]
\begin{picture}(270,90)(-55,0)

\multiput(0,30)(160,0){2}{\line(0,1){30}}
\multiput(0,30)(0,30){2}{\line(1,0){160}}

\multiput(40,30)(40,0){3}{\vector(0,-1){20}}
\multiput(10,80)(35,0){5}{\vector(0,-1){20}}

\put(68,41){SFA$_5$}

\put(6,84){$x_1$} \put(35,84){$u_1v_1$} \put(64,84){$u_1\vee v_1$}
\put(105,84){$u_2v_2$} \put(134,84){$u_2\vee v_2$}

\put(37,0){$c$} \put(71,0){$a_1b_1$} \put(104,0){$a_1\vee b_1$}

\put(220,45){\circle{10}} \put(260,45){\circle{10}}
\put(216.1,42.2){$\wedge$} \put(256.1,40.9){$\vee$}

\multiput(220,70)(40,0){2}{\vector(0,-1){20}}
\multiput(220,40)(40,0){2}{\vector(0,-1){20}}
\put(222.5,70){\vector(3,-2){33}}
\put(257.5,70){\vector(-3,-2){33}}

\put(216,74){$x_2$} \put(256,74){$x_3$} \put(211,10){$a_2b_2$}
\put(244,10){$a_2\vee b_2$}

\end{picture}
\caption{Block-diagram of the first CSA}
\end{figure}
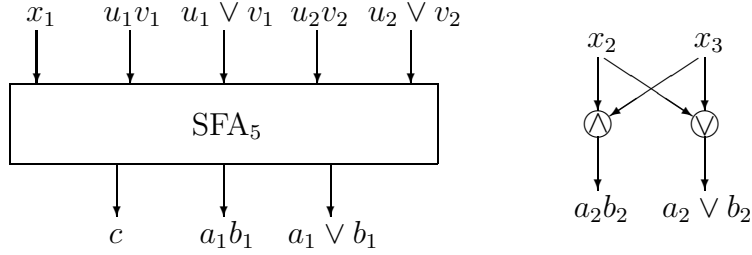

By analogy with the previous section for inputs $x_i$,
$(u_iv_i,\,u_i\vee v_i)$ and outputs $c$, $(a_ib_i,\,a_i\vee b_i)$
consider quantities $X_i$, $U_i$, $C$, $A_i$ respectively, which
correspond to the size of formulae implementing $x_i$, $u_iv_i$
(or $u_i\vee v_i$~--- here formulae for components in a pair have
equal size), $c$, $a_ib_i$.

According to~(\ref{esfa5}) the following inequalities hold:
\begin{equation}\label{esys5}
C \le 4X_1 + 8U_1 + 4U_2, \quad A_1 \le 2X_1 + 3U_1 + 2U_2, \quad
A_2 \le X_2 + X_3.
\end{equation}
One can easily check that the condition
\begin{equation*}
X_1^p + X_2^p + X_3^p - C^p > 0 \qquad U_1^p + U_2^p - A_1^p -
A_2^p > 0
\end{equation*}
is satisfied for $p=0.219978$, $X_1=1$, $X_2=X_3=0.031702$,
$U_1=1.018913$, $U_2=2$. As a consequence, $L_{B_0}(C_n) =
O(n^{4.546})$.

In the second example we encode triples of bits $u$, $v$, $w$ as
an ordered triple $\wt s = (s',\,s'',\,s''')$, supplemented with a
sum $s^{\oplus} = u \oplus v \oplus w$. Let us give formulae to
compute the code components:
$$ s' = \min \{ u,\,v,\,w \} = T_3^1(u,v,w) = u \vee v \vee w, $$
$$ s'' = T_3^2(u,v,w) = (u \vee v)w \vee uv, $$
$$ s''' = \max \{ u,\,v,\,w \} = T_3^3(u,v,w) = u v w. $$
Note that $s''$ and $s^{\oplus}$ are exact bits representing the
sum $u+v+w$.

CSA of the second example (see fig. 4) contains two $(7,4)$-CSA's
SFA$_7$ and SFA$'_7$, differing in encoding of a triple of inputs.

\begin{figure}[htb]
\begin{picture}(370,90)(-10,0)

\multiput(0,30)(160,0){2}{\line(0,1){30}}
\multiput(0,30)(0,30){2}{\line(1,0){160}}

\multiput(50,30)(60,0){2}{\vector(0,-1){20}}
\multiput(20,80)(60,0){3}{\vector(0,-1){20}}

\put(68,41){SFA$_7$}

\put(17,84){$x_1$} \put(66,84){$\wt s_1,\; s_1^{\oplus}$}
\put(126,84){$\wt s_2,\; s_2^{\oplus}$}

\put(47,0){$c_1$} \put(97,0){$\wt q_1,\; q_1^{\oplus}$}

\multiput(210,30)(160,0){2}{\line(0,1){30}}
\multiput(210,30)(0,30){2}{\line(1,0){160}}

\multiput(260,30)(60,0){2}{\vector(0,-1){20}}
\multiput(220,80)(30,0){4}{\vector(0,-1){20}}
\put(350,80){\vector(0,-1){20}}

\put(278,41){SFA$'_7$}

\put(216,84){$x_2$} \put(246,84){$x_3$} \put(276,84){$x_4$}
\put(306,84){$x_5$}

\put(336,84){$\wt s_3,\; s_3^{\oplus}$}

\put(257,0){$c_2$} \put(307,0){$\wt q_2,\; q_2^{\oplus}$}

\end{picture}
\caption{Block-diagram of the second CSA}
\end{figure}
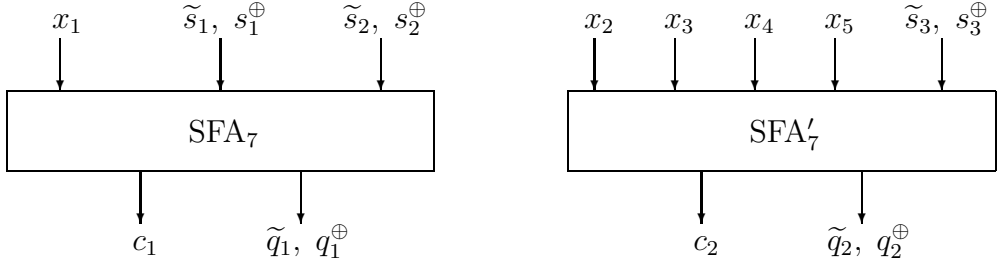

SFA$_7$ and SFA$'_7$ are functionally defined by equalities:
$$ c_1 + 2(q'_1 + q''_1 + q'''_1) = x_1 + s'_1 + s''_1 + s'''_1 + s'_2 + s''_2 + s'''_2, $$
$$ c_2 + 2(q'_2 + q''_2 + q'''_2) = x_2 + x_3 + x_4 + x_5 + s'_3 + s''_3 + s'''_3. $$
Outputs are implemented with formulae, structurally similar to
those for the $(7,3)$-CSA~\cite{ekh}:
\begin{equation}\label{esfa7}
\begin{array}{c}
c_1 = (x_1 \oplus s_1^{\oplus})\oplus s_2^{\oplus}  = s_2^{\oplus}
\overline{\left( x_1\overline{s_1^{\oplus}} \vee
\overline{x}_1s_1^{\oplus} \right)} \vee
 \overline{s_2^{\oplus}} \left( x_1\overline{s_1^{\oplus}} \vee \overline{x}_1s_1^{\oplus} \right), \\
c_2 = (x_2 \oplus x_3 \oplus x_4 \oplus x_5) \oplus s_3^{\oplus} =
\psi \overline{s_3^{\oplus}} \vee \overline{\psi} s_3^{\oplus}, \\
\phantom{\displaystyle \sum} \psi = (x_2\overline{x}_3 \vee
\overline{x}_2x_3)(x_4x_5 \vee \overline{x}_4\overline{x}_5) \vee
(x_2x_3 \vee \overline{x}_2\overline{x}_3)(x_4\overline{x}_5  \vee
\overline{x}_4x_5), \phantom{\displaystyle \sum} \\
\phantom{\displaystyle \sum}q'_1 = T_7^2(x_1,\wt s_1,\wt s_2) =
T_4^1(x_1, \wt s_1)s'_2
\vee T_4^2(x_1, \wt s_1) \vee s''_2, \phantom{\displaystyle \sum} \\
q''_1 = T_7^4(x_1,\wt s_1,\wt s_2) = \displaystyle
\bigvee_{i+j=4}T_4^i(x_1, \wt s_1)T_3^j(\wt s_2), \\
q^{\oplus}_1 = \overline{s'_2}\,T_4^2(x_1, \wt
s_1)\overline{T_4^4(x_1, \wt s_1)} \vee
s'_2\overline{s''_2}\,T_4^1(x_1, \wt s_1)\overline{T_4^3(x_1, \wt
s_1)} \; \vee \qquad \\ \qquad \vee \;
s''_2\overline{s'''_2}\left(T_4^4(x_1, \wt s_1) \vee
\overline{T_4^2(x_1, \wt s_1)}\right) \vee s'''_2\left(T_4^3(x_1,
\wt s_1) \vee \overline{T_4^1(x_1, \wt s_1)}\right).
 \end{array}
\end{equation}
Formula for $q'''_1 = T_7^6(x_1,\wt s_1,\wt s_2)$ is dual to that
for $q'_1$ up to substitution $s'_i$ by $s'''_i$. Formulae for
$\wt q_2$ and $q^{\oplus}_2$ coincide to those for $\wt q_1$ and
$q^{\oplus}_1$ up to implementation of the threshold funcion
$T_4(x_2,x_3,x_4,x_5)$.

Threshold functions $T_4^i(y_1,y_2,y_3,y_4) = T^i_4(y_1,\wt s)$
with three of variables allowing multiple encoding can be
implemented with formulae:
\begin{equation}\label{et4}
\begin{array}{c}
T_4^1 = y_1 \vee y_2 \vee y_3 \vee y_4 = y_1 \vee s', \\
T_4^2 = (y_1 \vee y_2)(y_3 \vee y_4) \vee y_1y_2 \vee y_3y_4 = y_1
s' \vee s''.
\end{array}
\end{equation}
Formulae for $T_4^3$ and $T_4^4$ are obtained in dual way with
substitution $s'$ by $s'''$.

According to the construction, formulae for $q'_1$ and $q'''_1$
($q'_2$ and $q'''_2$) have the equal size if the same holds for
inputs $s'_i$, $s'''_i$. Furthermore, the size of formula for
$q_i^{\oplus}$ is twice as much as that for $q''_i$.

As above, denote the complexity of formulae implementing inputs
$x_i$ and outputs $c_i$ by $X_i$ and $C_i$ respectively. For
inputs $\wt s_i$ consider the quantity $S_i=\max\{ S'_i,\,
S''_i/\alpha \}$, where $S'_i$ and $S''_i$ characterize the size
(to be more exact, an upper bound for the size) of formulae
implementing $s'_i$ and $s''_i$. Define quantities $Q_i$ for
outputs $\wt q_i$ analogously.

From (\ref{esfa7}) and (\ref{et4}) the inequalities follow:
\begin{equation}\label{esys7}
\begin{array}{c}
C_1 \le 4X_1 + 8\alpha S_1 + 4\alpha S_2, \\
\phantom{\displaystyle \sum}
C_2 \le 8(X_2+X_3+X_4+X_5) + 4\alpha S_3, \phantom{\displaystyle \sum} \\
Q_1 \le \max \left\{ 2X_1 + (\alpha+2)S_1 + (\alpha+1)S_2, \;
\frac4{\alpha}X_1 + \frac{2\alpha+4}{\alpha}S_1 +
\frac{\alpha+2}{\alpha}S_2 \right\}, \\ \phantom{\displaystyle
\sum} Q_2 \le \max \left\{ 3(X_2+X_3+X_4+X_5) + (\alpha+1)S_3, \;
\right. \qquad \qquad \quad \phantom{\displaystyle \sum}
\\ \qquad \qquad \qquad \qquad \qquad \qquad \qquad \left. \frac6{\alpha}(X_2+X_3+X_4+X_5) + \frac{\alpha+2}{\alpha}S_3
\right\}.
\end{array}
\end{equation}

With the use of~(\ref{esys7}) it can be checked that the condition
\begin{equation*}
X_1^p + X_2^p + X_3^p + X_4^p + X_5^p - C_1^p - C_2^p > 0 \qquad
S_1^p + S_2^p + S_3^p - Q_1^p - Q_2^p > 0
\end{equation*}
holds for $\alpha=1.6782$, $p=0.2204718$, $X_1=1$,
$X_2=X_3=X_4=X_5=0.3569540333$, $S_1=1.1282983248$,
$S_2=2.424317629$, $S_3=1.6884745179$. Therefore, $L_{B_0}(C_n) =
O(n^{4.5358})$.

The last bound probably can be improved even without constructing
CSA's more complicated than $SFA_5$ and $SFA_7$, if one uses all
three ways of encoding in unique CSA.

To estimate the complexity of an arbitrary symmetric function
consider a $(17,6)$-CSA with the standard encoding of inputs and
outputs, containing $SFA_7$ and pair of $SFA_5$'s in the bottom
and $(7,3)$-CSA similar to the CSA~\cite{ekh} at the top.
Non-standard outputs of SFA's are connected to the inputs of
$(7,3)$-CSA.\footnote{Vector of sizes of outputs of the CSA can be
produced from the vector of sizes of inputs via multiplication by
the matrix
$$ \left( \begin{matrix}
4 & 8 & 8 & 8 & 8 & 8 & 8 \\
0 & 0 & 0 & 0 & 0 & 0 & 0 \\
0 & 0 & 0 & 0 & 0 & 0 & 0   \\
12 & 16 & 16 & 24 & 24 & 24 & 24  \\
14 & 20 & 20 & 24 & 24 & 24 & 24  \\
 7 & 10 & 10 & 12 & 12 & 12 & 12
\end{matrix}\quad\begin{matrix}
 0 & 0 & 0 & 0 & 0 & 0 & 0 & 0 & 0 & 0 \\
 4 & 4 & 4 & 8 & 8 & 0 & 0 & 0 & 0 & 0 \\
 0 & 0 & 0 & 0 & 0 & 4 & 4 & 4 & 8 & 8  \\
 16 & 16 & 16 & 24 & 24 & 16 & 16 & 16 & 24 & 24 \\
 24 & 24 & 24 & 36 & 36 & 24 & 24 & 24 & 36 & 36 \\
 12 & 12 & 12 & 18 & 18 & 12 &12 & 12 & 18 & 18
\end{matrix}\right)$$ } One can verify that this CSA allows to implement
$C_n$ with complexity $O(n^{4.558})$ and a $k$-th significant bit
of $C_n$ with complexity $O(n^{3.8183}\cdot2^k)$. So, the bound
$L_{B_0}(S_n) = O(n^{4.8183})$ follows.

\section*{Appendix}

To make the presentation complete we provide here a method of
constructing formulae, which can be also found
in~\cite{eppz0,eppz,epz}.

\noindent {\bf 1. Implementation of $C_n$.}

Consider a CSA with inputs and outputs of $t$ types of encoding.
Let $x_{i,j}$ and $y_{i,j}$ denote inputs and outputs of $j$-th
type. Let the size $Y_{k,l}$ of the formula implementing an output
$y_{k,l}$ is a continuous, piecewise-linear and nondecreasing
(with respect to each argument) function of sizes $X_{i,j}$ of the
formulae implementing inputs $x_{i,j}$, where $X_{i,j}$, $Y_{k,l}$
take on arbitrary real non-negative values. Assume that if
$Y_{k,l}<X_{i,j}$ then $y_{k,l}$ does not depend on $x_{i,j}$. Let
the inequalities
\begin{equation}\label{ebal} \sum_i X_{i,j}^p - \sum_i Y_{i,j}^p
>0
\end{equation}
hold for some $p>0$, some $X_{i,j}>0$ and all $j=1,\ldots,t$.

We are to show how one can built a formula of size
$O\left(n^{1/p+o(1)}\right)$ to implement $C_n$.

Without loss of generality assume that $\min \{X_{i,j}\}=1< \min
\{Y_{i,j}\}$. As $Y$'s depend on $X$'s continuously there exists
such $\delta>0$ that for any $j$ inequality~(\ref{ebal}) remains
true after the substitution $X_{i,j}$ and $Y_{i,j}$ by parameters
$X'_{i,j} \in [X_{i,j}-\delta,\,X_{i,j}]$ and $Y'_{i,j} \in
[Y_{i,j},\,Y_{i,j}+\delta]$. Then there exist (small enough)
$\lambda>1$ and $d^X_{i,j},d^Y_{i,j} \in \mathbb Z$ such
that$\lambda^{d^X_{i,j}/p} \in [X_{i,j}-\delta,\,X_{i,j}]$ and
$\lambda^{d^Y_{i,j}/p} \in [Y_{i,j},\,Y_{i,j}+\delta]$ for all
$i$, $j$. Consequently for any $j$ the following inequality holds:
$$ \sum_i \lambda^{d^X_{i,j}}- \sum_i\lambda^{d^Y_{i,j}}>0. $$
Note that $\lambda^{d^X_{i,j}/p}$ is a lower bound for $X_{i,j}$
and $\lambda^{d^Y_{i,j}/p}$ is an upper bound for $Y_{i,j}$. Let
us name a number $d^X_{i,j}$ (respectively $d^Y_{i,j}$) level of
the input $x_{i,j}$ (output $y_{i,j}$). We can assume
$\min\{d^X_{i,j}\}=0$. Let $d=\max\{d^Y_{i,j}\}$.

Formula representing a bit of the function $C_n$ can be
constructed after the following pattern. The formula contains
CSA's on different levels. Each CSA can receive either inputs of
the formula, or outputs of other CSA's, or zero formulae as
inputs. CSA on a level $k$ receives inputs of $j$-th type on the
levels $d^X_{i,j}+k$ and produces outputs of the same type on the
levels $d^Y_{i,j}+k$. The formula receives its nonzero inputs
(i.e. symbols of variables) on the level $d$ and higher.

The formula is determined by the number $\left\lceil
cn\lambda^{-k} \right\rceil$ of CSA's on each level $k$, $0 \le k
\le \log_{\lambda} n$, where $c$ is a constant to be defined
later.

Let us estimate the number of inputs, including zeros, and the
number of outputs of a type $j$ in the formula. We will omit
indices $j$ in the argument below as it does not depend on $j$.

According to the construction, all outputs of the formula on the
levels $d$ and lower are zero. A total number of inputs (all zero)
on the same levels is $O(n)$. Difference between the number of
inputs and the number of outputs on a level $k$, $d \le k \le
\log_{\lambda} n$, is
\begin{multline*}
\sum_i \left\lceil cn\lambda^{d^X_i-k} \right\rceil - \sum_i
\left\lceil cn\lambda^{d^Y_i-k} \right\rceil =\\
= cn\lambda^{-k}\left( \sum_i \lambda^{d^X_i} - \sum_i
\lambda^{d^Y_i} \right) \pm O(1) =
\Theta\left(n\lambda^{-k}\right) \pm O(1).
\end{multline*}
On the levels higher than $\log_{\lambda} n$ the formula receives
and produces $O(1)$ inputs and outputs in total.

Hence, the formulae receives $\Theta(n)$ nonzero inputs and
produces $O(\log n)$ nonzero outputs (of $j$-th type). One can
choose $c$ large enough to provide not less than $n$ inputs for
any $j$.

Consider the size of outputs. It follows from the definition of
$\lambda$ that inputs and outputs on level $k$ are bounded above
by $\lambda^{k/p}$. Thus the size of outputs is
$\lambda^{(\log_{\lambda}n+O(1))/p} = O(n^{1/p})$.

To implement the $C_n$ function one has to take $\lfloor \log_2
n\rfloor+1$ parallel copies of the described pattern, zero some
inputs and re-commutate appropriately inputs and outputs on each
level. Final addition of $O(\log n)$ numbers can be implemented
with an arbitrary polynomial-size formula. So, the overall size of
the formulae for $C_n$ is $O\left(n^{1/p}\log^{O(1)}n\right)$.

\noindent {\bf 2. Formulae for symmetric functions.}

Consider a CSA with standard encoding of inputs and outputs. Let
$x_{s,i}$ and $y_{s,i}$ stand for inputs and outputs of $s$-th
significant bit, $s\ge0$. Let $X_{s,i}$ and $Y_{s,i}$ stand for
the size of corresponding formulae. For any $s$ define
$$  a_s = \sum_i X_{s,i}^p - \sum_i Y_{s,i}^p, $$
where we suppose sums over empty set of indices to be zero. Let
the inequalities
\begin{equation}\label{enu} a_0>0, \qquad \qquad \sum_s a_s \nu^{-s} > 0. \end{equation}
hold for some $p$, $X_{s,i}$ and $\nu\ge1$.

We will show that $l$-th significant bit of $C_n$ can be
implemented with a formula of size $O((\nu^ln)^{1/p+o(1)})$.

As above, choose an appropriate $\lambda>1$ and approximate
$X_{s,i}$ and $Y_{s,i}$ by integer powers of $\lambda$
preserving~(\ref{enu}) (denote the exponents by $d^X_{s,i}$,
$d^Y_{s,i}$). Without loss of generality assume
$\min\{d^X_{s,i}\}=0$. Define $d=\max\{d^Y_{s,i}\}$.

Let CSA on the level $k$ and of significance $l$ receive inputs of
significance $s+l$ on levels $d^X_{s,i}+k$ and produce outputs of
significance $s+l$ on levels $d^Y_{s,i}+k$. Consider a formula
containing $\lceil c\nu^ln\lambda^{-k} \rceil$ CSA's of
significance $l$, $0 \le l \le \log_2 n+1$, on a level $k$, $0 \le
k \le \log_{\lambda} (\nu^ln)$. Nonzero inputs of the formula are
received on the levels $d$ and higher, all of significance 0.

We are to estimate the number of inputs and outputs of
significance $l$ on level $k$. If $d \le k \le \log_{\lambda}
(\nu^ln)$, then difference between the number of inputs and the
number of outputs is
\begin{multline*}
\sum_{s,i} \left\lceil c\nu^{l-s}n\lambda^{d^X_{s,i}-k}
\right\rceil - \sum_{s,i}
\left\lceil c\nu^{l-s}n\lambda^{d^Y_{s,i}-k} \right\rceil =\\
= c\nu^ln\lambda^{-k}\sum_s a_s \nu^{-s} \pm O(1) =
\Theta(\nu^ln\lambda^{-k}) \pm O(1).
\end{multline*}
On the levels higher than $\log_{\lambda} (\nu^ln)$ the formula
receives and produces $O(1)$ inputs and outputs in total.

Therefore, the formula produces $O(\log n)$ outputs of any
significance. A choice of large enough constant $c$ provides at
least $n$ inputs of significance~0. Each output of significance
$l$ is implemented with a formula of size at most
$\lambda^{(\log_{\lambda}(\nu^ln)+O(1))/p} = O((\nu^ln)^{1/p})$.
Hence, $l$-th significant bit of $C_n$ can be implemented with a
formula of size $O((\nu^ln)^{1/p}\log^{O(1)}n)$.

Assuming $\nu\le2^p$ we obtain an upper bound $O(n^{1+1/p+o(1)})$
on the formula size complexity of the class $S_n$. The implied
formulae are constructed simply via representation of a symmetric
function as a function of the weight of its set of arguments and
decomposition along (new) variables.

\begin{thebibliography}{99}

\bibitem{elu}
Lupanov O. B. Asymptotic bounds for the complexity of control
systems. Moscow: MSU, 1984. 138 p. (in Russian)

\bibitem{ekh}
Khrapchenko V. M. The complexity of the realization of symmetrical
functions by formulae~// Mat. zametki. 1972, {\bf 11}(1), 109--120
(in Russian). [Engl. translation in Math. Notes Acad. Sci. USSR,
1972, {\bf 11}, 70--76.]

\bibitem{edkky}
Demenkov E., Kojevnikov A., Kulikov A., Yaroslavtsev G. New upper
bounds on the Boolean circuit complexity of symmetric functions~//
Inf. Proc. Letters. 2010, {\bf 110}(7), 264--267.

\bibitem{eju}
Jukna S. Boolean function complexity. Berlin, Heidelberg:
Springer-Verlag, 2012. 618 p.

\bibitem{eppz0}
Paterson M., Pippenger N., Zwick U. Faster circuits and shorter
formulae for multiple addition, multiplication and symmetric
Boolean functions~// Proc. 31st IEEE Symp. Found. Comput. Sci.,
1990, 642--650.

\bibitem{eppz}
Paterson M., Pippenger N., Zwick U. Optimal carry save networks~//
LMS Lecture Notes Series. {\bf 169}. Boolean function Complexity.
Cambridge University Press, 1992, 174--201.

\bibitem{epz}
Paterson M., Zwick U. Shallow circuits and concise formulae for
multiple addition and multiplication~// Comput. Complexity. 1993,
{\bf 3}, 262--291.

\bibitem{epe}
Peterson G. L. An upper bound on the size of formulae for
symmetric Boolean function. Tech. Report. 78--03--01. Univ.
Washington, 1978.

\bibitem{est}
Stockmeyer L. J. On the combinational complexity of certain
symmetric Boolean functions~// Math. Syst. Theory. 1977, {\bf 10},
323--336.

\end{thebibliography}
\end{document}

\title{Верхние оценки сложности формул для функции голосования
\footnote{Работа выполнена при финансовой поддержке РФФИ, проекты
11--01--00508 и 11--01--00792--а, и программы фундаментальных
исследований Отделения математических наук РАН <<Алгебраические и
комбинаторные методы математической кибернетики и информационные
системы нового поколения>> (проект <<Задачи оптимального синтеза
управляющих систем>>).}}
\date{}
\author{И. С. Сергеев}

\maketitle
\begin{abstract}
Показано, что сложность реализации функции подсчета числа единиц в
булевом наборе длины $n$ формулами над базисом двуместных булевых
функций не превосходит $O(n^{3,06})$, а над стандартным базисом
$O(n^{4,54})$. Как следствие, такие же оценки справедливы для
сложности любой пороговой симметрической функции $n$ переменных, в
частности, для функции голосования; сложность вычисления
произвольного разряда произведения $n$-разрядных двоичных чисел не
превосходит $O(n^{4,06})$ и $O(n^{5,54})$. Попутно получены оценки
сложности произвольных симметрических функций $n$ переменных:
$O(n^{3,23})$ и $O(n^{4,82})$ над теми же базисами.
\end{abstract}

\section{Введение}

В работе рассматривается сложность реализации симметрических
булевых функций формулами над базисом $B_2$ всех двуместных
булевых функций и над стандартным базисом $B_0 = \{ \wedge,
\vee,\neeg \}$. Понятия формулы и сложности $L_B(f)$ реализации
булевой функции $f$ формулами над базисом $B$ см. в~\cite{lu,ju}.

Рекордные на сегодня верхние оценки сложности и глубины реализации
симметрических функций как формулами, так и схемами из
функциональных элементов над полными базисами, опираются на
эффективную реализацию вектор-функции $C_n(x_1,\ldots,x_n)$
подсчета числа единиц в булевом наборе $(x_1,\ldots,x_n)$.
Сведение к вычислению функции $C_n$ используется при минимизации
глубины и сложности формул для умножения двоичных чисел.

В свою очередь, эффективные схемы и формулы для функции $C_n$
строятся из схем компрессоров (западный термин для схемы
компрессора --- CSA (carry save adder)). $(k,l)$-компрессор ширины
1 реализует булеву вектор-функцию $(x_1,\ldots,x_k) \to
(y_1,\ldots,y_l)$ с условием $\sum a_i x_i = \sum b_j x_j$, где
$k>l$, а константы $a_i, b_j$ являются целыми степенями двойки.
$(k,l)$-компрессор произвольной ширины строится из параллельных
копий компрессоров ширины 1 и позволяет сводить сложение $k$ чисел
к сложению $l$ чисел.

Используя подходящие компрессоры и метод~\cite{ppz}, в
работе~\cite{pz} получены оценки $L_{B_2}(C_n)=O(n^{3,13})$,
$L_{B_0}(C_n)=O(n^{4,57})$ уточняющие известные до этого оценки
%$L_{B_2}(C_n)=O(n^{3,21})$~\cite{ppz},
$L_{B_2}(C_n)=O(n^{3,32})$~\cite{pe} и
$L_{B_0}(C_n)=O(n^{4,62})$~\cite{kh} (о более ранних результатах
см. в~\cite{ppz}).

В недавней работе~\cite{dkky} построен компрессор (MDFA),
позволяющий реализовать функцию $C_n$ схемой над базисом $B_2$
рекордной сложности $4,5n$ (до этого была известна оценка
сложности $5n$). Эффективное использование компрессора~\cite{dkky}
предусматривает кодирование некоторых пар битов $u, v$ в виде
$(u\oplus v, v)$ (такой способ кодирования предложен в~\cite{st}).
Выигрыш в сложности проистекает из того, что (а) пара $(u\oplus v,
v)$ вычисляется не сложнее, чем пара $(u, v)$, (б) впоследствии
вместо бита $u$ используется сразу $u\oplus v$, что экономит одно
булево сложение.

Заметим, что в силу этих же причин следует ожидать, что MDFA
эффективен для построения формул. Действительно, компрессор
позволяет использовать на промежуточных этапах вычислений формулы
для $u \oplus v$, более короткие, чем суммарно формулы для $u$ и
$v$, минуя явное выполнение сложения $u$ и $v$.

Указанное наблюдение неявно уже использовалось при построении
рекордного для сложности формул компрессора в~\cite{pz}. Этот
компрессор содержит MDFA, к выходам которого присоединен
стандартный $(3,2)$-компрессор FA$_3$. Промежуточное кодирование
битов в виде $(u\oplus v, v)$~--- это ровно то, что делает
компрессор~\cite{pz} более эффективным, чем стандартный
$(3,2)$-компрессор (последний позволяет получить оценку \linebreak
$L_{B_2}(C_n)=O(n^{3,21})$, как показано в~\cite{ppz}).

Более того, по той причине, что MDFA экономит схемную сложность,
компрессор~\cite{pz} тоже обладает этим свойством: его схемная
сложность равна 14 (способ реализации функций в схеме несколько
отличается от способа их реализации в формуле) поэтому он
позволяет реализовать функцию $C_n$ со сложностью $(14/3)n$. Таким
образом, барьер $5n$ мог быть преодолен еще в 90-е гг.

Естественно ожидать, что компрессор~\cite{pz} не является
оптимальным способом использования MDFA для построения коротких
формул. Далее мы покажем, что используя MDFA более независимым
образом, можно получить оценку $L_{B_2}(C_n)=O(n^{3,06})$.
Предлагаемый способ, однако, тоже не использует возможности MDFA
оптимально.

Аналогичная идея может быть реализована и в случае стандартного
базиса. В этом случае применяется кодирование пар битов $u$ и $v$
упорядоченными парами $(uv,\;u\vee v)$. Выходы компрессора в
старших разрядах в такой кодировке являются монотонными функциями
входов (причем пороговыми, если все входы относятся к одному
разряду). Это может давать преимущество, поскольку в известных
конструкциях компрессоров над базисом $B_0$ наиболее
сложнореализуемыми являются немонотонные выходы.

Ниже мы построим $(5,3)$-компрессор SFA$_5$~--- аналог MDFA для
монотонной кодировки битов. Используя его, совсем просто получить
оценку $L_{B_0}(C_n)=O(n^{4,55})$. Чуть лучшая оценка
$L_{B_0}(C_n)=O(n^{4,54})$ получается при помощи более сложной
конструкции, основанной на $(7,3)$-компрессоре из~\cite{kh} и
монотонной кодировке троек битов.

Новые оценки сложности для функции голосования не приводят к
автоматическому улучшению оценок сложности реализации формулами
класса $S_n$ симметрических функций $n$ переменных, поскольку
известные методы~\cite{kh,ppz0} ограничивают эффективность
компрессоров со входами и выходами в различных кодировках.

Впрочем, из нескольких экземпляров компрессоров с измененной
кодировкой битов несложно построить компрессоры со стандартной
кодировкой, при помощи которых методом~\cite{ppz0} получаются
оценки $L_{B_2}(S_n)=O(n^{3,23})$, $L_{B_0}(S_n)=O(n^{4,82})$,
уточняющие известные ранее оценки \linebreak
$L_{B_2}(S_n)=O(n^{3,30})$~\cite{ppz0},
$L_{B_2}(S_n)=O(n^{3,37})$~\cite{pe},
$L_{B_0}(S_n)=O(n^{4,85})$~\cite{ppz0},
$L_{B_0}(S_n)=O(n^{4,93})$~\cite{kh}.

%Применяя метод~\cite{ppz0}\footnote{Метод~\cite{ppz0} основан на
%аккуратном воплощении идеи~\cite{kh} реализации младших разрядов
%функции $C_n$ формулами существенно меньшей сложности, чем для
%старших разрядов.}, можно получить оценки
% и

Подчеркнем, что конструкции элементарных компрессоров,
используемые в настоящей работе, принципиально такие же или, во
всяком случае, не более сложные, чем конструкции из предшествующих
работ. Новые оценки сложности получены исключительно в результате
эксплуатации идеи альтернативной кодировки битов.

\section{Конструкция для базиса $B_2$}

\begin{figure}[htb]
\begin{picture}(170,90)(-115,0)

\multiput(0,30)(160,0){2}{\line(0,1){30}}
\multiput(0,30)(0,30){2}{\line(1,0){160}}

\multiput(40,30)(40,0){3}{\vector(0,-1){20}}
\multiput(10,80)(35,0){5}{\vector(0,-1){20}}

\put(64,41){MDFA}

\put(7,84){$x$} \put(41,84){$v_1$} \put(63,84){$u_1\oplus v_1$}
\put(111,84){$v_2$} \put(132,84){$u_2\oplus v_2$}

\put(37,0){$c$} \put(77,0){$b$} \put(107,0){$a\oplus b$}

\end{picture}
\caption{Блок-схема MDFA}
\end{figure}

На рис. 1 изображена блок-схема компрессора MDFA, а ниже приведено
определение функционирования компрессора и формулы, по которым
вычисляются функции на его выходах.

$$ 2(a+b)+c = x+u_1+u_2+v_1+v_2. $$
\begin{equation}\label{mdfa}
\begin{array}{c}
\phantom{\displaystyle \sum}  c  = x \oplus (u_1 \oplus v_1)
\oplus (u_2 \oplus v_2), \qquad
    b  = (x\oplus v_1)(u_1 \oplus v_1) \oplus v_1, \phantom{\displaystyle \sum}  \\
  (a \oplus b) = ((x\oplus v_1) \vee (u_1 \oplus v_1)) \oplus
    (x\oplus (u_1 \oplus v_1) \oplus v_2)\overline{(u_2 \oplus
    v_2)}.
\end{array}
\end{equation}

Для получения оценки сложности $O(n^{3,06})$ мы используем
компрессор, изображенный на рис. 2. С точностью до кодировки
выходов, он состоит из двух изолированных компрессоров MDFA.

\begin{figure}[htb]
\begin{picture}(370,120)(-13,30)

%%% MDFA carcas

\put(0,30){
\begin{picture}(160,90) % MDFA
\multiput(0,30)(160,0){2}{\line(0,1){30}}
\multiput(0,30)(0,30){2}{\line(1,0){160}}
\multiput(40,30)(40,0){3}{\vector(0,-1){20}}
\put(10,80){\vector(0,-1){20}} \put(45,110){\vector(0,-1){50}}
\put(46.7,110){\vector(4,-3){29}} \put(80,110){\vector(0,-1){20}}
\put(80,85){\circle{10}} \put(80,80){\line(0,1){10}}
\put(75,85){\line(1,0){10}}
\multiput(80,80)(35,0){3}{\vector(0,-1){20}} \put(64,41){MDFA}
\end{picture}}

%%%  signatures

\put(11,114){$x_1$} \put(44,144){$x_2$} \put(80,144){$x_3$}
\put(115,114){$v_1$} \put(137,114){$u_1 \oplus v_1$}

\put(40,30){$c_1$} \put(81,30){$b_1$} \put(108,30){$a_1\oplus
b_1$}

%%% MDFA carcas

\put(195,45){
\begin{picture}(160,90) % MDFA
\multiput(0,30)(160,0){2}{\line(0,1){30}}
\multiput(0,30)(0,30){2}{\line(1,0){160}}
\multiput(40,30)(40,0){3}{\vector(0,-1){20}}
\multiput(10,80)(35,0){5}{\vector(0,-1){20}} \put(64,41){MDFA}
\end{picture}}

%%%  signatures

\put(204,129){$x_4$} \put(239,129){$v_2$} \put(262,129){$u_2
\oplus v_2$} \put(309,129){$v_3$} \put(332,129){$u_3 \oplus v_3$}

\put(234,45){$c_2$} \put(275,45){$b_2$} \put(302,45){$a_2\oplus
b_2$}

\end{picture}
\caption{Конструкция основного компрессора}
\end{figure}

Компрессор имеет входы и выходы двух типов: одиночные биты и пары
битов в кодировке $(u\oplus v,v)$.

Входам и выходам первого типа поставим в соответствие сложность
реализующих их формул, обозначим ее через $X_i$ для входов $x_i$ и
через $C_i$~--- для выходов $c_i$. Входам и выходам второго типа в
кодировке $(u\oplus v,v)$ поставим в соответствие величину $\max
\{ V,\, U^+/\alpha\}$, где $V$, $U^+$~--- сложность (или верхняя
оценка сложности) формул, реализующих соответственно $v$ и
$u\oplus v$, а $\alpha$~--- параметр, который будет выбран
позднее. Указанную величину для входов $(u_i\oplus v_i,v_i)$ и
выходов $(a_i\oplus b_i,b_i)$ обозначим через $U_i$ и $A_i$
соответственно.

Согласно формулам~(\ref{mdfa}), справедливы соотношения:
\begin{equation}\label{sys}
\begin{array}{c}
C_1 \le X_1 + X_2 + X_3 + \alpha U_1, \\ \phantom{\displaystyle
\sum}
 C_2 \le X_4 + \alpha U_2 + \alpha U_3, \phantom{\displaystyle \sum}
 \\ \phantom{\displaystyle
\sum}
 A_1 \le \max \left\{ X_1 + X_2 + 3X_3, \;
\frac2{\alpha}X_1 + \frac2{\alpha}X_2 + \frac3{\alpha}X_3 +
\frac{\alpha+1}{\alpha}U_1 \right\}, \phantom{\displaystyle \sum}
\\ \phantom{\displaystyle \sum}
 A_2 \le \max \left\{ X_4 + (\alpha+2)U_2, \;
\frac2{\alpha}X_4 + \frac{2\alpha+1}{\alpha}U_2 +
\frac{\alpha+1}{\alpha}U_3 \right\}. \phantom{\displaystyle \sum}
\end{array}
\end{equation}

Из~\cite{ppz,pz} следует, что если при некотором $p>0$, некоторых
$X_i > 0$ и $U_i > 0$ (а также при некотором $\alpha>0$)
справедливо
\begin{equation}\label{inq}
\begin{array}{c}
X_1^p + X_2^p + X_3^p + X_4^p - C_1^p - C_2^p > 0, \\
\phantom{\displaystyle \sum}U_1^p + U_2^p + U_3^p - A_1^p - A_2^p
> 0, \phantom{\displaystyle \sum}
\end{array}
\end{equation}
то $L_{B_2}(C_n) = O\left(n^{1/p+o(1)}\right)$, см. также
Приложение. Используя вместо величин $C_i$ и $A_i$ верхние оценки
из~(\ref{sys}), можно проверить, что неравенства~(\ref{inq})
справедливы при $\alpha=2,906$, $p=0,327781$, $X_1=X_2=1$,
$X_3=0,5149081$, $X_4=1,9198088$, $U_1=1,2176395$,
$U_2=1,0031176$, $U_3=2,3573055$. Следовательно, $L_{B_2}(C_n) =
O(n^{3,0509})$.

Чтобы получить более точные оценки эффективности MDFA, можно
разбить второй тип входов и выходов на несколько типов с
различными значениями $\alpha$ (например, равномерно
распределенными на некотором отрезке) и рассмотреть набор из
MDFA-компрессоров со входами и выходами всевозможных типов.
Впрочем, без привлечения дополнительных соображений это достаточно
трудоемкая вычислительная задача. В малой части указанное
соображение учтено в конструкции компрессора рис.~2: отношение
размеров формул для $x_2\oplus x_3$ и $x_2$ отличается от
$\alpha$.

%Пусть $k'=0,7885\log_2 n$. Применяя метод~\cite{ppz0} и используя
%$(6,3)$-комп\-рес\-сор~\cite{pz}, можно проверить, что если $k\le
%k'$, то $k$-й разряд функции $C_n$ реализуется со сложностью
%$O(n^{3,0509}2^{k-k'})$. Следовательно, $L_{B_2}(S_n) =
%O(n^{3,2625})$.

%Эту оценку можно дальше улучшать, используя более эффективные
%компрессоры со входами и выходами одного типа.
Чтобы оценить сложность реализации произвольной симметрической
функции, рассмотрим последовательность компрессоров со стандартной
кодировкой битов, в которой компрессор с номером $m$ состоит из
$m$ штук последовательно соединенных MDFA и внешнего компрессора
$FA_3$ (выходы каждого MDFA присоединяются ко входам $v_2$,
$u_2\oplus v_2$ следующего MDFA в обозначениях рис.~1). Первый
член этой последовательности ($m=1$) совпадает с компрессором
из~\cite{pz}. При $m=4$ получаем $(15,6)$-компрессор, который,
взятый отдельно, позволяет реализовать $C_n$ формулой сложности
$O(n^{3,089})$. Вектор сложности выходов этого компрессора
получается из вектора сложности входов умножением на матрицу
$$ \left( \begin{matrix}
0 & 0 & 0 & 0 & 0  \\
0 & 0 & 0 & 0 & 0  \\
0 & 0 & 0 & 0 & 1    \\
0 & 1 & 1 & 1 & 2   \\
1 & 2 & 2 & 3 & 3   \\
1 & 4 & 4 & 9 & 3
\end{matrix}\quad\begin{matrix}
 0 & 0 & 0 & 0 & 0 & 1 & 1 & 1 & 1 & 1 \\
 0 & 0 & 1 & 1 & 1 & 1 & 2 & 2 & 2 & 3 \\
 1 & 1 & 2 & 2 & 3 & 1 & 2 & 3 & 3 & 6  \\
 2 & 3 & 3 & 3 & 6 & 1 & 2 & 3 & 3 & 6 \\
 3 & 6 & 3 & 3 & 6 & 1 & 2 & 3 & 3 & 6 \\
 3 & 6 & 3 & 3 & 6 & 1 & 2 & 3 & 3 & 6
\end{matrix}\right)$$

Применяя этот $(15,6)$-компрессор способом~\cite{ppz0}, $k$-й
разряд функции $C_n$ можно реализовать со сложностью
$O(n^{2,2285}\cdot2^k)$, откуда получаем $L_{B_2}(S_n) =
O(n^{3,2285})$, обоснование см. в Приложении.

\section{Конструкция для базиса $B_0$}

В этом параграфе мы приведем два примера компрессоров, эффективных
для построения формул для $C_n$ над базисом $B_0$. Первый
компрессор изображен на рис. 3. Функционирование компрессора
SFA$_5$, лежащего в основе данного примера ~--- такое же, как у
MDFA. Выходы вычисляются по формулам:
\begin{equation}\label{sfa5}
\begin{array}{c}
\phantom{\displaystyle \sum} c = (x_1 \oplus (u_1 \oplus v_1))
\oplus (u_2 \oplus v_2) =
 \psi \overline{\chi} \vee  \overline{\psi} \chi, \phantom{\displaystyle \sum} \\
%\phantom{\displaystyle \sum}
\psi =  x_1\left((u_1v_1) \vee \overline{(u_1 \vee v_1)}\right)
\vee \overline{x_1}\overline{(u_1v_1)}(u_1 \vee
 v_1), \;\,  \chi = \overline{(u_2v_2)}(u_2 \vee  v_2), %\phantom{\displaystyle \sum}
 \\
a_1b_1 = T_5^4(x_1,u_1,v_1,u_2,v_2) = \displaystyle\bigvee_{i+j=4}
T_3^i(x_1,u_1,v_1) T_2^j(u_2,v_2) =
\\ = (x_1(u_1 \vee v_1) \vee (u_1v_1))(u_2v_2) \vee x_1(u_1v_1)(u_2 \vee
v_2), \end{array}
\end{equation}
где через $T_n^k$ обозначается монотонная пороговая функция $n$
переменных с порогом $k$; выход $a_1 \vee b_1  =
T_5^2(x_1,u_1,v_1,u_2,v_2)$ является функцией, двойственной к
$a_1b_1$, поэтому реализуется двойственной формулой.

\begin{figure}[htb]
\begin{picture}(270,90)(-55,0)

\multiput(0,30)(160,0){2}{\line(0,1){30}}
\multiput(0,30)(0,30){2}{\line(1,0){160}}

\multiput(40,30)(40,0){3}{\vector(0,-1){20}}
\multiput(10,80)(35,0){5}{\vector(0,-1){20}}

\put(68,41){SFA$_5$}

\put(6,84){$x_1$} \put(35,84){$u_1v_1$} \put(64,84){$u_1\vee v_1$}
\put(105,84){$u_2v_2$} \put(134,84){$u_2\vee v_2$}

\put(37,0){$c$} \put(71,0){$a_1b_1$} \put(104,0){$a_1\vee b_1$}

\put(220,45){\circle{10}} \put(260,45){\circle{10}}
\put(216.1,42.2){$\wedge$} \put(256.1,40.9){$\vee$}

\multiput(220,70)(40,0){2}{\vector(0,-1){20}}
\multiput(220,40)(40,0){2}{\vector(0,-1){20}}
\put(222.5,70){\vector(3,-2){33}}
\put(257.5,70){\vector(-3,-2){33}}

\put(216,74){$x_2$} \put(256,74){$x_3$} \put(211,10){$a_2b_2$}
\put(244,10){$a_2\vee b_2$}

\end{picture}
\caption{Блок-схема компрессора из первого примера}
\end{figure}

По аналогии с предыдущим параграфом входам $x_i$,
$(u_iv_i,\,u_i\vee v_i)$ и выходам $c$, $(a_ib_i,\,a_i\vee b_i)$
поставим в соответствие величины $X_i$, $U_i$, $C$, $A_i$,
характеризующие сложность формул, реализующих $x_i$, $u_iv_i$ (или
$u_i\vee v_i$~--- в рассматриваемом примере формулы для элементов
пары имеют одинаковый размер), $c$, $a_ib_i$.

Из формул~(\ref{sfa5}) вытекают соотношения:
\begin{equation}\label{sys5}
C \le 4X_1 + 8U_1 + 4U_2, \quad A_1 \le 2X_1 + 3U_1 + 2U_2, \quad
A_2 \le X_2 + X_3.
\end{equation}
Как можно проверить исходя из этих неравенств, условие
\begin{equation*}
X_1^p + X_2^p + X_3^p - C^p > 0 \qquad U_1^p + U_2^p - A_1^p -
A_2^p > 0
\end{equation*}
выполнено при $p=0,219978$, $X_1=1$, $X_2=X_3=0,031702$,
$U_1=1,018913$, $U_2=2$. Следовательно, $L_{B_0}(C_n) =
O(n^{4,546})$.

Во втором примере мы используем кодирование троек битов $u$, $v$,
$w$ в виде упорядоченного набора $\wt s = (s',\,s'',\,s''')$,
дополненного суммой $s^{\oplus} = u \oplus v \oplus w$. Приведем
формулы для вычисления компонент кода:
$$ s' = \min \{ u,\,v,\,w \} = T_3^1(u,v,w) = u \vee v \vee w, $$
$$ s'' = T_3^2(u,v,w) = (u \vee v)w \vee uv, $$
$$ s''' = \max \{ u,\,v,\,w \} = T_3^3(u,v,w) = u v w. $$
Отметим, что $s''$ и $s^{\oplus}$ являются в точности битами суммы
$u+v+w$.

Компрессор из второго примера (см. рис. 4) состоит из двух
$(7,4)$-компрессоров SFA$_7$ и SFA$'_7$, отличающихся только
кодировкой одной из групп входов.

\begin{figure}[htb]
\begin{picture}(370,90)(-10,0)

\multiput(0,30)(160,0){2}{\line(0,1){30}}
\multiput(0,30)(0,30){2}{\line(1,0){160}}

\multiput(50,30)(60,0){2}{\vector(0,-1){20}}
\multiput(20,80)(60,0){3}{\vector(0,-1){20}}

\put(68,41){SFA$_7$}

\put(17,84){$x_1$} \put(66,84){$\wt s_1,\; s_1^{\oplus}$}
\put(126,84){$\wt s_2,\; s_2^{\oplus}$}

\put(47,0){$c_1$} \put(97,0){$\wt q_1,\; q_1^{\oplus}$}

\multiput(210,30)(160,0){2}{\line(0,1){30}}
\multiput(210,30)(0,30){2}{\line(1,0){160}}

\multiput(260,30)(60,0){2}{\vector(0,-1){20}}
\multiput(220,80)(30,0){4}{\vector(0,-1){20}}
\put(350,80){\vector(0,-1){20}}

\put(278,41){SFA$'_7$}

\put(216,84){$x_2$} \put(246,84){$x_3$} \put(276,84){$x_4$}
\put(306,84){$x_5$}

\put(336,84){$\wt s_3,\; s_3^{\oplus}$}

\put(257,0){$c_2$} \put(307,0){$\wt q_2,\; q_2^{\oplus}$}

\end{picture}
\caption{Блок-схема компрессора из второго примера}
\end{figure}

Функционирование компрессоров SFA$_7$ и SFA$'_7$ определяют
соотношения:
$$ c_1 + 2(q'_1 + q''_1 + q'''_1) = x_1 + s'_1 + s''_1 + s'''_1 + s'_2 + s''_2 + s'''_2, $$
$$ c_2 + 2(q'_2 + q''_2 + q'''_2) = x_2 + x_3 + x_4 + x_5 + s'_3 + s''_3 + s'''_3. $$
Выходы вычисляются по формулам, структурно таким же, как формулы
для $(7,3)$-компрессора~\cite{kh}:
\begin{equation}\label{sfa7}
\begin{array}{c}
c_1 = (x_1 \oplus s_1^{\oplus})\oplus s_2^{\oplus}  = s_2^{\oplus}
\overline{\left( x_1\overline{s_1^{\oplus}} \vee
\overline{x}_1s_1^{\oplus} \right)} \vee
 \overline{s_2^{\oplus}} \left( x_1\overline{s_1^{\oplus}} \vee \overline{x}_1s_1^{\oplus} \right), \\
c_2 = (x_2 \oplus x_3 \oplus x_4 \oplus x_5) \oplus s_3^{\oplus} =
\psi \overline{s_3^{\oplus}} \vee \overline{\psi} s_3^{\oplus}, \\
\phantom{\displaystyle \sum} \psi = (x_2\overline{x}_3 \vee
\overline{x}_2x_3)(x_4x_5 \vee \overline{x}_4\overline{x}_5) \vee
(x_2x_3 \vee \overline{x}_2\overline{x}_3)(x_4\overline{x}_5  \vee
\overline{x}_4x_5), \phantom{\displaystyle \sum} \\
\phantom{\displaystyle \sum}q'_1 = T_7^2(x_1,\wt s_1,\wt s_2) =
T_4^1(x_1, \wt s_1)s'_2
\vee T_4^2(x_1, \wt s_1) \vee s''_2, \phantom{\displaystyle \sum} \\
q''_1 = T_7^4(x_1,\wt s_1,\wt s_2) = \displaystyle
\bigvee_{i+j=4}T_4^i(x_1, \wt s_1)T_3^j(\wt s_2), \\
q^{\oplus}_1 = \overline{s'_2}\,T_4^2(x_1, \wt
s_1)\overline{T_4^4(x_1, \wt s_1)} \vee
s'_2\overline{s''_2}\,T_4^1(x_1, \wt s_1)\overline{T_4^3(x_1, \wt
s_1)} \; \vee \qquad \\ \qquad \vee \;
s''_2\overline{s'''_2}\left(T_4^4(x_1, \wt s_1) \vee
\overline{T_4^2(x_1, \wt s_1)}\right) \vee s'''_2\left(T_4^3(x_1,
\wt s_1) \vee \overline{T_4^1(x_1, \wt s_1)}\right).
 \end{array}
\end{equation}
Формула для $q'''_1 = T_7^6(x_1,\wt s_1,\wt s_2)$ получается
двойственным образом по отношению к формуле для $q'_1$ с заменой
$s'_i$ на $s'''_i$. Формулы для $\wt q_2$ и $q^{\oplus}_2$
устроены аналогично формулам для $\wt q_1$ и $q^{\oplus}_1$,
отличаясь только реализацией пороговых функций
$T_4(x_2,x_3,x_4,x_5)$.

Пороговые функции $T_4^i(y_1,y_2,y_3,y_4) = T^i_4(y_1,\wt s)$
четырех переменных, три из которых могут задаваться в разных
кодировках, реализуются формулами:
\begin{equation}\label{t4}
\begin{array}{c}
T_4^1 = y_1 \vee y_2 \vee y_3 \vee y_4 = y_1 \vee s', \\
T_4^2 = (y_1 \vee y_2)(y_3 \vee y_4) \vee y_1y_2 \vee y_3y_4 = y_1
s' \vee s''.
\end{array}
\end{equation}
Формулы для $T_4^3$ и $T_4^4$ получаются двойственным образом с
заменой $s'$ на $s'''$.

По построению, формулы для $q'_1$ и $q'''_1$ ($q'_2$ и $q'''_2$)
имеют одинаковую сложность при условии, что то же верно для входов
$s'_i$, $s'''_i$. Кроме того, сложность формулы для $q_i^{\oplus}$
вдвое больше, чем сложность формулы для $q''_i$.

Как и выше, будем обозначать через $X_i$ и $C_i$ сложность формул,
реализующих входы $x_i$ и выходы $c_i$ соответственно. Входам $\wt
s_i$ поставим в соответствие величину $S_i=\max\{ S'_i,\,
S''_i/\alpha \}$, где величины $S'_i$ и $S''_i$ характеризуют
сложность (точнее, верхнюю оценку сложности) формул, реализующих
$s'_i$ и $s''_i$. Аналогично определим величины $Q_i$,
соответствующие выходам $\wt q_i$.

Из (\ref{sfa7}) с учетом (\ref{t4}) следуют соотношения:
\begin{equation}\label{sys7}
\begin{array}{c}
C_1 \le 4X_1 + 8\alpha S_1 + 4\alpha S_2, \\
\phantom{\displaystyle \sum}
C_2 \le 8(X_2+X_3+X_4+X_5) + 4\alpha S_3, \phantom{\displaystyle \sum} \\
Q_1 \le \max \left\{ 2X_1 + (\alpha+2)S_1 + (\alpha+1)S_2, \;
\frac4{\alpha}X_1 + \frac{2\alpha+4}{\alpha}S_1 +
\frac{\alpha+2}{\alpha}S_2 \right\}, \\ \phantom{\displaystyle
\sum} Q_2 \le \max \left\{ 3(X_2+X_3+X_4+X_5) + (\alpha+1)S_3, \;
\right. \qquad \qquad \quad \phantom{\displaystyle \sum}
\\ \qquad \qquad \qquad \qquad \qquad \qquad \qquad \left. \frac6{\alpha}(X_2+X_3+X_4+X_5) + \frac{\alpha+2}{\alpha}S_3
\right\}.
\end{array}
\end{equation}

Используя~(\ref{sys7}), можно проверить, что условие
\begin{equation*}
X_1^p + X_2^p + X_3^p + X_4^p + X_5^p - C_1^p - C_2^p > 0 \qquad
S_1^p + S_2^p + S_3^p - Q_1^p - Q_2^p > 0
\end{equation*}
выполнено при $\alpha=1,6782$, $p=0,2204718$, $X_1=1$,
$X_2=X_3=X_4=X_5=0,3569540333$, $S_1=1,1282983248$,
$S_2=2,424317629$, $S_3=1,6884745179$. Следовательно,
$L_{B_0}(C_n) = O(n^{4,5358})$.

Вероятно, эту оценку можно улучшить даже без привлечения более
сложных, чем $SFA_5$ и $SFA_7$, конструкций, если использовать все
три способа кодировки битов в одном компрессоре.

Для оценки сложности произвольной симметрической функции
рассмотрим $(17,6)$-компрессор со стандартной кодировкой битов,
составленный из компрессора $SFA_7$ и двух компрессоров $SFA_5$ в
основании, упорядоченные тройка и две пары выходов которых
присоединены ко входам $(7,3)$-компрессора, который реализуется в
стиле~\cite{kh}. Вектор сложности выходов этого компрессора
получается из вектора сложности входов умножением на матрицу
$$ \left( \begin{matrix}
4 & 8 & 8 & 8 & 8 & 8 & 8 \\
0 & 0 & 0 & 0 & 0 & 0 & 0 \\
0 & 0 & 0 & 0 & 0 & 0 & 0   \\
12 & 16 & 16 & 24 & 24 & 24 & 24  \\
14 & 20 & 20 & 24 & 24 & 24 & 24  \\
 7 & 10 & 10 & 12 & 12 & 12 & 12
\end{matrix}\quad\begin{matrix}
 0 & 0 & 0 & 0 & 0 & 0 & 0 & 0 & 0 & 0 \\
 4 & 4 & 4 & 8 & 8 & 0 & 0 & 0 & 0 & 0 \\
 0 & 0 & 0 & 0 & 0 & 4 & 4 & 4 & 8 & 8  \\
 16 & 16 & 16 & 24 & 24 & 16 & 16 & 16 & 24 & 24 \\
 24 & 24 & 24 & 36 & 36 & 24 & 24 & 24 & 36 & 36 \\
 12 & 12 & 12 & 18 & 18 & 12 &12 & 12 & 18 & 18
\end{matrix}\right)$$

Можно проверить, что такой компрессор позволяет реализовать
функцию $C_n$ со сложностью $O(n^{4,558})$ и отдельно $k$-й разряд
этой функции~--- со сложностью $O(n^{3,8183}\cdot2^k)$, откуда
получаем $L_{B_0}(S_n) = O(n^{4,8183})$.

%\section{Заключение}

\section*{Приложение}

В этом разделе для полноты изложения мы приводим метод построения
формул. Его можно также найти в~\cite{ppz0,ppz,pz}.

\noindent {\bf 1. Реализация функции $C_n$%при помощи компрессоров с различными типами входов и выходов
.}

Пусть входы и выходы некоторого компрессора относятся к одному из
$t$ типов. Обозначим через $x_{i,j}$ входы, а через $y_{i,j}$~---
выходы $j$-го типа. Пусть размер $Y_{k,l}$ формулы, реализующей
выход $y_{k,l}$, является непрерывной кусочно-линейной однородной
и неубывающей по каждому аргументу функцией размеров $X_{i,j}$
формул, реализующих входы $x_{i,j}$, где величины $X_{i,j}$,
$Y_{k,l}$ могут принимать произвольные неотрицательные
вещественные значения. При этом если $Y_{k,l}<X_{i,j}$, то
$y_{k,l}$ не зависит от $x_{i,j}$. Пусть при некотором $p>0$ и
некоторых $X_{i,j}>0$ при любом $j=1,\ldots,t$ выполняется
\begin{equation}\label{bal}
\sum_i X_{i,j}^p - \sum_i Y_{i,j}^p >0.
\end{equation}
Покажем, как построить формулу для $C_n$ сложности
$O\left(n^{1/p+o(1)}\right)$.

Без ограничения общности можем считать, что $\min \{X_{i,j}\}=1<
\min \{Y_{i,j}\}$. В силу непрерывной зависимости $Y_{i,j}$ от
$X_{i,j}$ найдется такое $\delta>0$, что при любом $j$
неравенство~(\ref{bal}) остается справедливым при подстановке
в~(\ref{bal}) параметров $X'_{i,j} \in [X_{i,j}-\delta,\,X_{i,j}]$
и $Y'_{i,j} \in [Y_{i,j},\,Y_{i,j}+\delta]$ вместо соответствующих
$X_{i,j}$ и $Y_{i,j}$. Тогда найдется (достаточно малое)
$\lambda>1$, для которого существуют $d^X_{i,j},d^Y_{i,j} \in
\mathbb Z$ такие, что $\lambda^{d^X_{i,j}/p} \in
[X_{i,j}-\delta,\,X_{i,j}]$ и $\lambda^{d^Y_{i,j}/p} \in
[Y_{i,j},\,Y_{i,j}+\delta]$ для всех $i$, $j$. Следовательно, при
любом $j$ имеем
$$ \sum_i \lambda^{d^X_{i,j}}- \sum_i\lambda^{d^Y_{i,j}}>0, $$
причем $\lambda^{d^X_{i,j}/p}$ является оценкой снизу для
$X_{i,j}$, а $\lambda^{d^Y_{i,j}/p}$ является оценкой сверху для
$Y_{i,j}$. Назовем число $d^X_{i,j}$ (соответственно $d^Y_{i,j}$)
уровнем входа $x_{i,j}$ (выхода $y_{i,j}$). Мы можем полагать, что
$\min\{d^X_{i,j}\}=0$. Пусть $d=\max\{d^Y_{i,j}\}$.

Формулу, реализующую произвольный разряд функции $C_n$, построим
по следующему шаблону. Формула строится из компрессоров,
расположенных на различных уровнях: входами компрессоров могут
быть входы формулы, выходы компрессоров, расположенных на уровнях
с меньшими номерами, а также тождественно нулевые формулы.
Компрессор, расположенный на уровне $k$, принимает входы типа $j$
на уровнях $d^X_{i,j}+k$ и производит выходы соответствующего типа
на уровнях $d^Y_{i,j}+k$. Ненулевые входы формулы (т.е. символы
переменных) располагаются на уровнях $d$ и выше.

На каждом уровне $k$ формулы, $0 \le k \le \log_{\lambda} n$,
располагается $\left\lceil cn\lambda^{-k} \right\rceil$
компрессоров, где $c$~--- некоторая константа, которая будет
определена позднее.

Оценим число входов, включая тождественно нулевые, и число выходов
каждого типа $j$ в схеме. Поскольку следующее рассуждение не
зависит от $j$, индекс $j$ в нем опустим.

По построению, все выходы схемы на уровнях меньше $d$ являются
нулевыми. Суммарное число входов (все они нулевые) на тех же
уровнях есть $O(n)$. На уровне $k$, $d \le k \le \log_{\lambda}
n$, разность между числом входов и числом выходов есть
\begin{multline*}
\sum_i \left\lceil cn\lambda^{d^X_i-k} \right\rceil - \sum_i
\left\lceil cn\lambda^{d^Y_i-k} \right\rceil =\\
= cn\lambda^{-k}\left( \sum_i \lambda^{d^X_i} - \sum_i
\lambda^{d^Y_i} \right) \pm O(1) =
\Theta\left(n\lambda^{-k}\right) \pm O(1).
\end{multline*}
На уровнях выше $\log_{\lambda} n$ схема суммарно принимает и
производит $O(1)$ входов и выходов.

Таким образом, схема имеет $\Theta(n)$ ненулевых входов и
производит $O(\log n)$ ненулевых выходов ($j$-го типа). Выбирая
достаточно большое $c$, можно добиться, чтобы число входов было не
меньше $n$ при любом~$j$.

Оценим размер выходов. Согласно выбору $\lambda$, размер формул на
уро\-в\-не~$k$ оценивается сверху как $\lambda^{k/p}$.
Следовательно, формулы, реализующие выходы, имеют сложность не
выше $\lambda^{(\log_{\lambda}n+O(1))/p} = O(n^{1/p})$.

Функция $C_n$ реализуется посредством $\lfloor \log_2 n\rfloor+1$
параллельных копий описанного шаблона, в которых некоторые входы
обнуляются, и выполняется подходящая перекоммутация входов и
выходов на каждом уровне. Заключительное сложение $O(\log n)$
чисел можно реализовать произвольной формулой полиномиальной
сложности, поэтому окончательно сложность вычисления $C_n$
оценивается как $O\left(n^{1/p}\log^{O(1)}n\right)$.

\noindent {\bf 2. Реализация симметрических функций%отдельных разрядов функции $C_n$ при помощи компрессоров со входами и выходами одного типа
.}

Рассмотрим компрессор, имеющий входы и выходы одного типа. Через
$x_{s,i}$ и $y_{s,i}$ обозначим входы и выходы в $s$-м разряде,
$s\ge0$, через $X_{s,i}$ и $Y_{s,i}$ обозначим размер
соответствующих формул.
%Пусть зависимость $Y_{s,i}$ от совокупности $X_{s,i}$ выражается кусочно-линейной, однородной,
%непрерывной и неубывающей по каждому аргументу функцией.
При любом $s$ положим
$$  a_s = \sum_i X_{s,i}^p - \sum_i Y_{s,i}^p, $$
где суммы по пустому множеству индексов считаются равными нулю.
Пусть при некоторых $p$, $X_{s,i}$ и $\nu\ge1$ выполнены
неравенства
\begin{equation}\label{nu} a_0>0, \qquad \qquad \sum_s a_s \nu^{-s} > 0. \end{equation}
Покажем, что $l$-й разряд функции $C_n$ можно реализовать формулой
сложности $O((\nu^ln)^{1/p+o(1)})$.

Как и в предыдущем пункте выберем подходящее приближение для
$X_{s,i}$ и $Y_{s,i}$ в виде целых степеней $\lambda>1$,
сохраняющее неравенства~(\ref{nu}) (показатели степеней обозначим
соответственно через $d^X_{s,i}$, $d^Y_{s,i}$). Без ограничения
общности считаем, что $\min\{d^X_{s,i}\}=0$. Обозначим
$d=\max\{d^Y_{s,i}\}$.

Пусть компрессор разряда $l$ и уровня $k$ принимает входы разряда
$s+l$ на уровнях $d^X_{s,i}+k$ и производит выходы разряда $s+l$
на уровнях $d^Y_{s,i}+k$. Формулу построим по схеме, в которой при
любом $l$, $0 \le l \le \log_2 n+1$, на уровне $k$, $0 \le k \le
\log_{\lambda} (\nu^ln)$, размещено $\lceil c\nu^ln\lambda^{-k}
\rceil$ компрессоров. Ненулевые входы формулы располагаются в
нулевом разряде на уровнях $d$ и выше.

Оценим число входов и выходов, относящихся к уровню $k$ и
разряду~$l$. Если $d \le k \le \log_{\lambda} (\nu^ln)$, то
разность между числом входов и числом выходов есть
\begin{multline*}
\sum_{s,i} \left\lceil c\nu^{l-s}n\lambda^{d^X_{s,i}-k}
\right\rceil - \sum_{s,i}
\left\lceil c\nu^{l-s}n\lambda^{d^Y_{s,i}-k} \right\rceil =\\
= c\nu^ln\lambda^{-k}\sum_s a_s \nu^{-s} \pm O(1) =
\Theta(\nu^ln\lambda^{-k}) \pm O(1).
\end{multline*}
На уровнях выше $\log_{\lambda} (\nu^ln)$ схема суммарно принимает
и производит $O(1)$ входов и выходов.

Следовательно, схема в каждом разряде производит $O(\log n)$
выходов. Подходящий выбор константы $c$ обеспечивает не менее $n$
входов в нулевом разряде. Формулы, реализующие выходы в разряде
$l$, имеют сложность не выше
$\lambda^{(\log_{\lambda}(\nu^ln)+O(1))/p} = O((\nu^ln)^{1/p})$.
Как следствие, $l$-й разряд функции $C_n$ реализуется формулой
сложности $O((\nu^ln)^{1/p}\log^{O(1)}n)$.

Если $\nu\le2^p$, то для реализации класса $S_n$ формулами
получаем оценку сложности $O(n^{1+1/p+o(1)})$: симметрическая
функция представляется как функция суммы своих аргументов~--- для
этой функции формула строится методом последовательного разложения
по (новым) переменным.